# Revealing Phosphorus Nitrides up to the Megabar Regime: Synthesis of α′-P$_3$N$_5$, δ-P$_3$N$_5$ and PN$_2$

Dr. Dominique Laniel,[1,2]* Dr. Florian Trybel,[3] Dr. Adrien Néri,[4] Yuqing Yin,[1,5] Andrey Aslandukov,[1,4] Dr. Timofey Fedotenko,[6] Saiana Khandarkhaeva,[4] Dr. Ferenc Tasnádi,[3] Dr. Stella Chariton,[7] Dr. Carlotta Giacobbe,[8] Dr. Eleanor Lawrence Bright,[8] Dr. Michael Hanfland,[8] Dr. Vitali Prakapenka,[7] Prof. Dr. Wolfgang Schnick,[9] Prof. Dr. Igor A. Abrikosov,[3] Prof. Dr. Leonid Dubrovinsky,[4] Prof. Dr. Natalia Dubrovinskaia[1,3]

**Affiliations:**

[1]Material Physics and Technology at Extreme Conditions, Laboratory of Crystallography, University of Bayreuth, 95440 Bayreuth, Germany

[2]Centre for Science at Extreme Conditions and School of Physics and Astronomy, University of Edinburgh, EH9 3FD Edinburgh, United Kingdom

[3]Department of Physics, Chemistry and Biology (IFM), Linköping University, SE-581 83, Linköping, Sweden

[4]Bayerisches Geoinstitut, University of Bayreuth, 95440 Bayreuth, Germany

[5]State Key Laboratory of Crystal Materials, Shandong University, Jinan 250100, China

[6]Deutsches Elektronen-Synchrotron, Notkestr. 85, 22607 Hamburg, Germany

[7]Center for Advanced Radiation Sources, University of Chicago, Chicago, Illinois 60637, United States

[8]European Synchrotron Radiation Facility, B.P.220, F-38043 Grenoble Cedex, France

[9]Department of Chemistry, University of Munich (LMU), Butenandtstrasse 5-13, 81377 Munich, Germany

*Correspondence to dominique.laniel@ed.ac.uk

**Abstract**

Non-metal nitrides are an exciting field of chemistry, featuring a significant number of compounds that can possess outstanding material properties. This characteristic relies on maximizing the number of strong covalent bonds, with crosslinked XN$_6$ octahedra frameworks being particularly intriguing. In this study, the phosphorus-nitrogen system was studied up to 137 GPa in laser-heated diamond anvil cells and three previously unobserved phases were synthesized and characterized by single-crystal X-ray diffraction, Raman spectroscopy measurements and density functional theory calculations. δ-P$_3$N$_5$ and PN$_2$ were found to form at 72 and 134 GPa, respectively, and both feature dense 3D networks of the so far elusive PN$_6$ units. The two are ultra-incompressible, having a bulk modulus of K$_0$ = 322 GPa for δ-P$_3$N$_5$ and of K$_0$ = 339 GPa for PN$_2$. Upon decompression below 7 GPa, δ-P$_3$N$_5$ undergoes a transformation into a novel α′-P$_3$N$_5$ solid, stable at ambient conditions, that has a unique structure type based on PN$_4$ tetrahedra. The formation of α′-P$_3$N$_5$ underlines that a phase space otherwise inaccessible can be explored through high-pressure formed phases.

## Introduction

Non-metal nitrides are prone to form dense and highly crosslinked networks made up of strongly bound alternating X and N atoms (X being a non-metal). These solids are highly sought-after for their exceptional materials properties, such as incompressibility and mechanical hardness, combined with a tendency of having high thermal stability, photocatalytic activity, chemical inertness and a wide bandgap with optoelectrical properties.[1–7] Prominent examples of binary non-metal nitrides are the BN polymorphs (c-BN, h-BN),[4,8] $Si_3N_4$,[2,9] and diverse triazine- and heptazine-based carbon nitrides.[6,10–12] The structural chemistry of these compounds enables their impressive elastic properties through compact polyhedral arrangements, typically $XN_4$ tetrahedra with strong covalent X-N bonds. For non-metal elements capable of forming $XN_5$ and $XN_6$ polyhedra under high pressure, these can result in a significant hike in incompressibility, as demonstrated in the case of $\beta$-$Si_3N_4$ ($K_0$ = ~237 GPa)[13] → $\gamma$-$Si_3N_4$ ($K_0$ = ~300 GPa)[9] and expected for $\alpha$-$P_3N_5$ ($K_0$ = 134 GPa)[14] → $V_3O_5$-type $P_3N_5$ ($K_0$ = 303 GPa).[14] At very high pressure, this can also be accompanied by a simultaneous change of stoichiometry and the formation of pernitride $N_2^{x-}$ units, such as in $\gamma$-$Si_3N_4$ → $SiN_2$,[15] $\beta$-$Ge_3N_4$ → $GeN_2$,[15] *etc.* and remarkably, these pressure-formed higher coordinated compounds are often found recoverable to ambient conditions.

While transformation from $XN_4$ to $XN_6$ units was already observed in pressure-formed binary group 14 element nitrides ($SiN_2$, $GeN_2$, $SnN_2$)[15] and a chalcogen nitride ($SN_2$),[16] $XN_6$ octahedra are still eluding binary pnictogen nitrides.[17–19] This gap in our empirical understanding of non-metal nitrides is especially striking based on the significant efforts that were devoted to their synthesis, particularly in binary phosphorus nitrides. Still, the many experimental studies,[18–21] supported by theoretical calculations,[14,22,23] amounted to the synthesis of three $P_3N_5$ polymorphs, namely $\alpha$- and $\beta$-$P_3N_5$, both formed at near ambient conditions,[18,20] and $\gamma$-$P_3N_5$ produced at 11 GPa and 1500 K.[19] The exact crystal structure of $\beta$-$P_3N_5$ is still unknown.[18,20] $\alpha$-$P_3N_5$ can be represented by the Niggli formula[24] $_\infty^3[P_3^{[4]}N_3^{[2]}N_2^{[3]}]$—where the coordination number of a given atom is provided in superscripted square brackets (*i.e.* $P_3^{[4]}$ represents three phosphorus atoms each fourfold coordinated) and the dimensionality of the network is given by the number of dimensions, in superscript, in which the structural unit has an infinite extension (*i.e.* for the infinite polymeric 3D $\alpha$-$P_3N_5$, it is $_\infty^3[...]$). $\alpha$-$P_3N_5$ features corner- and edge-sharing $PN_4$ units, while the pressure-formed $\gamma$-$P_3N_5$ contains both $PN_4$ and $PN_5$ polyhedra according to $_\infty^3[P_1^{[4]}P_2^{[5]}N_1^{[2]}N_4^{[3]}]$. All these phosphorus nitrides are composed solely of heteroatomic P-N bonds. Further endeavours to produce a binary phosphorus nitride with $PN_6$ octahedra by compressing $\gamma$-$P_3N_5$ to 80 GPa did not result in any phase transformation, although laser-heating $\gamma$-$P_3N_5$ to 2000 K between 67 and 70 GPa resulted in the appearance of new Raman modes originating from a compound with an unsolved structure.[21] Outside the substance class of nitrides, the existence of $PN_6$ units has been confirmed in the molecular hexaazidophosphate anion.[25]

Not before recently, the existence of $PN_6$ units has been confirmed in pressure-synthesized ternary phosphorus nitrides[26] $\beta$-$BP_3N_6$[27] and spinel-type $BeP_2N_4$.[28] Moreover, theoretical calculations support the formation of phosphorus nitrides with $PN_6$ octahedra, with kyanite[23] and $V_3O_5$[22] as structure candidates being most likely. Importantly, these calculations also reveal that compositions other than $P_3N_5$ might be stable under pressure, namely $PN_2$ and $PN_3$. Both of the latter are expected to contain the elusive $PN_6$ units as well.[22] As such, there are substantial reasons to believe that further experiments on the P-N system could contribute to completing the more than 25-year hunt for binary pnictogen nitride solid with $XN_6$ octahedra, finally getting back in line with group 14 and 16 (chalcogen) nitrides.

**Results and Discussion**

Here, we report the synthesis of three new phosphorus nitrides, namely α′-$P_3N_5$, δ-$P_3N_5$ and $PN_2$, obtained through direct nitridation of elemental phosphorus. δ-$P_3N_5$ and $PN_2$ both feature the hitherto elusive $PN_6$ octahedra. To accomplish this, three BX90-type diamond anvil cells (DAC)[29] were prepared to investigate the high-pressure behavior of phosphorus-nitrogen samples up to 137 GPa. As described in detail in the Supporting Information, one DAC was loaded with red phosphorus and the two others with the black phosphorus allotrope. In either case the samples of phosphorus were loaded along with gaseous molecular nitrogen. At 72 GPa, laser-heating a sample to temperatures above 2600 K resulted in the formation of a new solid, based on the appearance of new diffraction lines which could not be explained by the known phosphorus nitride γ-$P_3N_5$,[21] nitrogen (ε-$N_2$, ζ-$N_2$ or ι-$N_2$)[30–32] or phosphorus (phase III) (Figure S1).[33,34] Single-crystal X-ray diffraction (SC-XRD) of selected areas of the polycrystalline sample was performed and the collected data enabled a full structural solution (see Table S1 for the crystallographic details). The formed compound has a monoclinic unit cell (space group $C2/c$, no. 15) with lattice parameters $a = 8.418(5)$, $b = 4.325(6)$ and $c = 8.418(5)$ Å and $β = 110.96(7)°$, $Z = 4$, $V = 205.4(4)$ Å$^3$. It is composed of five crystallographically unique atoms, P1 and P2, which are on the 8$f$ and 4$a$ Wyckoff sites, and N1, N2 and N3 on the 4$e$, 8$f$ and 8$f$ Wyckoff positions, respectively. Having the $P_3N_5$ stoichiometry, this novel solid is hereafter named δ-$P_3N_5$. δ-$P_3N_5$ has the high-temperature $V_3O_5$ structure type[35] and was previously predicted from theoretical calculations.[14,22] As shown in Figure 1, the two distinct P atoms are sixfold coordinated by nitrogen atoms, forming the sought after $PN_6$ octahedra. The N1 and N2 atoms are both fourfold coordinated by phosphorus atoms and thus exhibit an ammonium type character, while the N3 atom is threefold coordinated by P atoms. The atoms' environment can be expressed by the Niggli formula $^3_\infty[P_3^{[6]}N_2^{[3]}N_3^{[4]}]$. δ-$P_3N_5$ is the first phosphorus nitride featuring nitrogen atoms with ammonium type $N^{[4]}$ character. However, the latter are known in γ-$Si_3N_4$[9,36] and in nitridosilicates.[37,38] As expected, the average P-$N^{[4]}$ distance (1.694(9) to 1.895(6) Å) is significantly longer than its P-$N^{[3]}$ (1.600(6) to 1.644(5) Å) counterpart. The average of all P-N bonds in δ-$P_3N_5$ is 1.712(6) Å in length, comparable to those found in γ-$P_3N_5$ at 1 bar (average bond length of 1.69 Å)[19]. δ-$P_3N_5$ was also synthesized at 118 GPa from laser-heating red phosphorus and molecular nitrogen (Table S2).

The experimentally determined structural model of δ-$P_3N_5$ was perfectly reproduced by Kohn-Sham density functional theory (DFT) calculations (Table S1), and phonon calculations found the structure dynamically stable at 72 GPa (Figure S2). The computed electronic bands and projected electron density of states (p-eDOS) of δ-$P_3N_5$ show that it is a semiconductor with an indirect bandgap of 2.8 eV (Figure S3) calculated using the Perdew-Burke-Ernzerhof (PBE) approximation to exchange and correlation which will most like underestimate the real band gap. The calculated electron localization function (ELF) (Figure S4) displays the polar-covalent nature of the bonding between P and N. As expected, the analysis of ELF isosurfaces showed that both $N^{[4]}$ centers do not display a lone electron pair, donated to the N-P dative bond (Figure S4). On the opposite, the N3 atom ($N^{[3]}$ center), not expected to form a dative bond, shows a lone electron pair. Dative bonding can be considered as a natural response to a pressure increase, serving to extend both nitrogen's coordination and the compound's density.

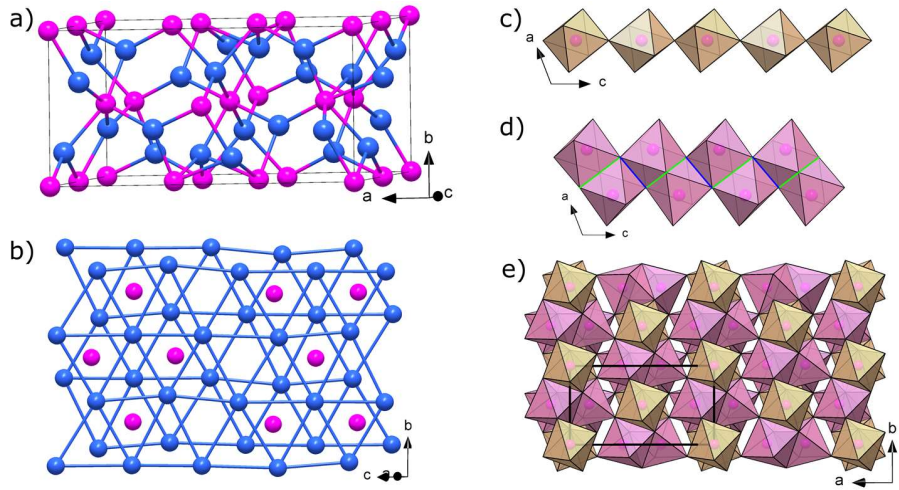

Figure 1: Structure of δ-$P_3N_5$ at 72 GPa. a) The unit cell. b) The distorted *hcp* arrangement of the N atoms with P atoms filling 3/5 of the octahedral voids. c) Chains formed by the P2-centered $PN_6$ octahedra. d) Anatase ($TiO_2$) like chains formed by the P1-centered $PN_6$ octahedra. The green and blue lines indicate face- and edge-sharing $PN_6$ units. e) Four unit cells with polyhedra drawn. The pink and blue spheres represent P and N atoms, respectively.

The structural arrangement of δ-$P_3N_5$ can be described as a distorted hexagonal closest packing (*hcp*) of N atoms with the 3/5 of octahedral voids filled with P atoms. This is similar to β-$BP_3N_6$ which also features a nitrogen distorted *hcp* arrangement with P atoms in the octahedral voids, but additionally with B atoms in the tetrahedral cavities.[27] The structure of δ-$P_3N_5$ can also be broken down into two interconnected chainlike structural elements running along the [001] direction, as illustrated in Figure 1, with P atoms on $y \approx 0$ and $y \approx 1/2$. The first chain is composed of 'double octahedra', *i.e.* two face-sharing octahedra with the P1 atoms at their centers (P1-P1 distance of 2.407(5) Å at 72 GPa) that are linked together by their edges to form a chain. Exhibiting face-sharing octahedra is an indication of the (polar) covalent bonding in δ-$P_3N_5$; as known from Pauling's third rule,[39] this configuration is very unfavorable for cations in ionic structures. The other type of chain, comprised of P2-centered $PN_6$ units, is composed of single octahedra linked by corner sharing. The two sets of chains are further crosslinked by edge- and corner-sharing $PN_6$ units, and alternate in both [100] and [010] directions.

Raman measurements on δ-$P_3N_5$ were performed at 82 GPa and vibrations at frequencies of 416, 626, 673 and 838 cm$^{-1}$ were detected. As shown in Figure S5, these modes match very well with those previously observed from an unidentified reaction product from a γ-$P_3N_5$ sample laser-heated between 67 and 70 GPa, thereby strongly suggesting it to be δ-$P_3N_5$ as well.[21]

δ-$P_3N_5$ produced at 72 GPa was characterized by SC-XRD during its decompression. The full crystallographic data at 118 GPa is shown in Table S2, while the decompression data, with the unit cell parameters of δ-$P_3N_5$ extracted, are shown in Figure 2 as well as in Table S4. The δ-$P_3N_5$ polymorph was detected down to a minimum pressure of 7 GPa, and experimentally determined to have a bulk modulus of $K_0$ = 322(6) GPa ($V_0$ = 245.7(6) Å$^3$, $K'$ = 4 (fixed); outlier point at 72 GPa not included in the fit). The incompressibility value, being above 300 GPa, qualifies δ-$P_3N_5$ as an ultra-incompressible solid similar with spinel-type $BeP_2N_4$.[28] Its incompressibility is vastly greater than that of the lower pressure polymorph γ-$P_3N_5$, determined to have a $K_0$ = 130.27(43) GPa ($K'$ = 4 (fixed)). This difference can easily be explained by the polymorphs' respective crystal chemistry, *i.e.* γ-$P_3N_5$ being composed of a mixture of $PN_4$ and $PN_5$ units while δ-$P_3N_5$ is exclusively made up of $PN_6$ octahedra, even including face-sharing.

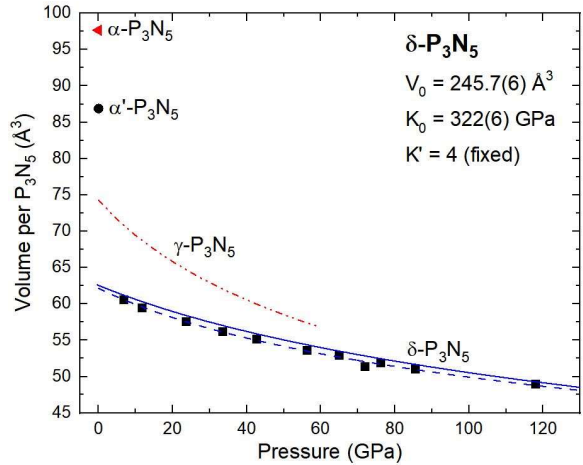

Figure 2: Volume per $P_3N_5$ units with respect to pressure for all known polymorphs of $P_3N_5$. The black squares and circle represent experimental datapoints on δ-$P_3N_5$ and α′-$P_3N_5$ obtained in our experiments. The black dashed line is a fit of the δ-$P_3N_5$ experimental data with a second order Birch-Murnaghan equation of state, with parameters of $V_0$ = 245.7(6) Å$^3$, $K_0$ = 322(6) GPa and K′ fixed to a value of 4. The full blue line represents the calculated equation of state for δ-$P_3N_5$. The red dashed and dotted lines and the red triangle represent the literature experimental data on γ-$P_3N_5$[21] and α-$P_3N_5$,[18] respectively. β-$P_3N_5$ is not included as its exact structure is not known, although it is likely to be a stacking variant of α-$P_3N_5$.[18]

Subsequently, the DAC containing δ-$P_3N_5$ was fully opened, thereby releasing nitrogen gas and exposing the sample to air. Despite a thorough X-ray diffraction mapping of the sample chamber, δ-$P_3N_5$ could no longer be observed. However, diffraction lines from yet another novel solid were identified, and its crystal structure was solved from the collected single-crystal data. This revealed the formation of a fifth $P_3N_5$ polymorph, hereafter named α′-$P_3N_5$, which is very likely to be the phase transformation product of δ-$P_3N_5$ at pressures below 7 GPa. Similar phase transformations of high-pressure phases upon full pressure release have been observed in a number of systems, including $BeN_4$[40] and $Mg_2N_4$.[41] The formation of α′-$P_3N_5$ further emphasizes that compounds formed at high pressures can serve as precursors to explore a phase space otherwise inaccessible.

At ambient conditions, α′-$P_3N_5$ was found to have a monoclinic unit cell (space group $P2_1/c$, no. 14) with lattice parameters $a$ = 9.2557(6), $b$ = 4.6892(3), $c$ = 8.2674(6) Å and $β$ = 104.160(6)°, Z = 4, V = 347.92(5) Å$^3$. All eight crystallographically distinct atoms, three P and five N, are occupying 4$e$ Wyckoff sites. The full crystallographic data is available in Table S3 and reciprocal space unwarps shown in Figure S6. As shown in Figure 3, the structure is composed of cross-linked $PN_4$ tetrahedra and has the Niggli formula $^3_\infty\left[P_3^{[4]}N_3^{[2]}N_2^{[3]}\right]$—the same as the known ambient condition polymorph α-$P_3N_5$ (see Figure 3),[18] and similar to β-$Si_3N_4$ $^3_\infty\left[Si_3^{[4]}N_4^{[3]}\right]$. When viewed along the $b$-axis, it can be understood as two types of $PN_4$ layers (light green and light pink tetrahedra) joined together by connecting corner-sharing $PN_4$ units (light orange tetrahedra). The light green tetrahedra, centered by P3, are forming distorted rings composed of six tetrahedra in the $bc$-plane, four corner-sharing and two edge-sharing—an arrangement of tetrahedra which had previously been observed in HP-$NiB_2O_4$.[42] The light pink tetrahedra (P1 atom at their center) are producing a corner-sharing single zweier chain (according to F. Liebau's[43] nomenclature) running along the $b$-axis. While (non-condensed) zweier chains also occur in $Mg_2PN_3$ and $Ca_2PN_3$,[44,45] the crystal structure of α′-$P_3N_5$ is, to the best of our knowledge, a new structure type. At 1 bar, the P-N bond lengths vary between 1.51(1) Å and 1.78(1) Å, similar to those in α-$P_3N_5$ (1.51 to 1.74 Å).[18] Still, α′-$P_3N_5$ is found to be 12.4% denser than α-$P_3N_5$, as it can be deducted from Figure 2. This can be explained by the tighter packing of phosphorus atoms in α′-$P_3N_5$, with an average P-P distance of 2.900(5) Å compared to 2.98 Å in α-$P_3N_5$.[18]

DFT calculations on α′-$P_3N_5$ confirm the experimental structural model (Table S3) and demonstrate its dynamical stability (Figure S7). Akin to δ-$P_3N_5$, the P-N bonds in α′-$P_3N_5$ are seen from ELF calculations to be polar covalent (Figure S8). α′-$P_3N_5$ is found to be a wide band gap semiconductor, having a calculated direct band gap of 3.45 eV (using PBE, Figure S9), which is significantly lower than the calculated indirect bandgap of 5.21 eV for α-$P_3N_5$.[7] Despite α′-$P_3N_5$ and α-$P_3N_5$ featuring many similarities regarding their crystal chemistry, as exemplified by their identical Niggli formula, the much smaller band gap of α′-$P_3N_5$ can be hypothesized to be a consequence of its higher density, leading to a more thorough overlap of the electronic orbitals. The bulk modulus of α′-$P_3N_5$ is calculated to be $K_0$ = 95.4 GPa ($K'$ = 3.89), which is in line with that of α-$P_3N_5$ ($K_0$ = 87 GPa ($K'$ = 2.0) or 99 GPa ($K'$ = 1.9), depending on the calculations.[23]

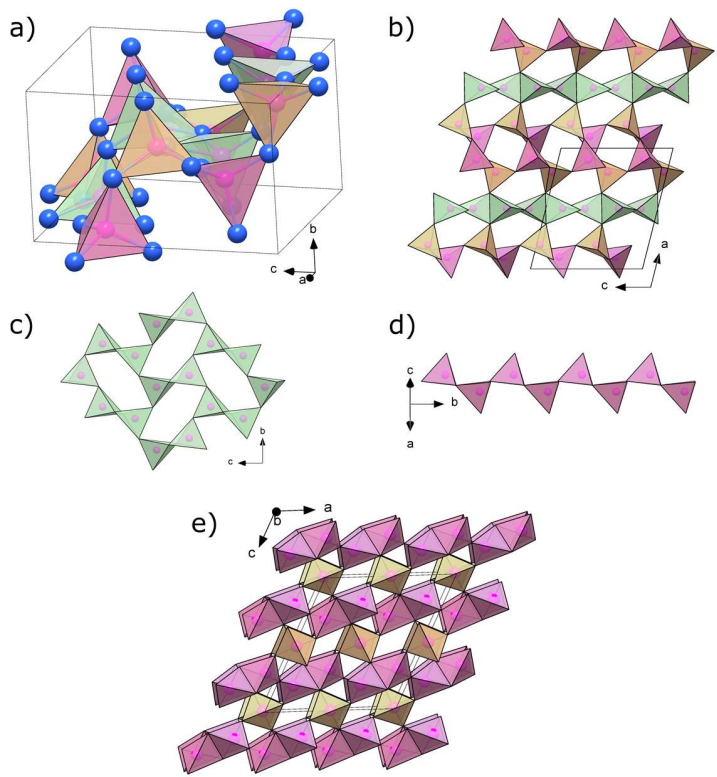

Figure 3: Structure of α′-$P_3N_5$ at 1 bar. Blue and pink spheres represent N and P atoms, respectively. The light pink, light orange and light green $PN_4$ tetrahedra are centered by P1, P2 and P3, respectively. a) Unit cell. b) Multiple unit cells viewed along the *b*-axis, allowing to see the stacking of the different tetrahedra along the *a*-axis. c) Slice of the *bc*-plane, where the arrangement of the light green tetrahedra is visible. Four distorted rings, each composed of six tetrahedra are seen. Each ring is composed of four corner-sharing and two edge-sharing $PN_4$. Such an arrangement of tetrahedra has previously been observed in HP-$NiB_2O_4$.[42] d) Single zweier chain of light pink tetrahedra running along the *b*-axis. e) For comparison, the structure of α-$P_3N_5$ at 1 bar viewed along the [010] direction.[18] $PN_4$ polyhedra forming zweier single chains are drawn in light pink, and are alternately corner- and edge-sharing. The chains are connected together by additional $PN_4$ tetrahedra, drawn in light orange.

Further experiments on the phosphorus-nitrogen system were also conducted above 120 GPa. Sample laser-heating to temperatures above 2000 K at 134 GPa resulted in the formation of yet another binary phosphorus nitride with unprecedented $PN_2$ stoichiometry. Its crystal structure was solved from SC-XRD data at both 134 GPa as well as at 137 GPa after further compression (see Tables S5 and S6; see Figure S10 for reciprocal space unwarps). $PN_2$ has a simple pyrite-type structure (cubic, space group *Pa*-3, no. 205)—akin to group 14 element nitrides SiN$_2$, GeN$_2$ and SnN$_2$[15] as well as the platinum group metal nitrides PtN$_2$ and PdN$_2$[46–49] and has a unit cell parameter $a$ = 4.0127(14) Å with Z = 4 and $V$ = 64.61(4) Å$^3$) at 134 GPa. As shown in Figure 4, it is composed of two chemically and crystallographically distinct atoms, P1 and N1, respectively on Wyckoff positions 4$a$ and 8$c$, respectively. PN$_2$ is a 3D polymeric compound that can be expressed by the Niggli formula $^3_\infty\left[P_1^{[6]}N_2^{[4]}\right]$ and is composed of corner-sharing PN$_6$ octahedra cross-linked together through nitrogen dimers. Each nitrogen atom is therefore forming three P-N bonds (1.699(3) Å, 134 GPa) and a single N-N bond (1.418(6) Å, 134 GPa). The P-N bond length is marginally shorter than the average P-N contacts in δ-P$_3$N$_5$ at 72 GPa (1.712(6) Å), while the length of the nitrogen dimer strongly suggests a single-bond, $i.e.$ a pernitride unit ([N$_2$]$^{4-}$).[50] From the Niggli formula of the PN$_2$ compound, the N$^{[4]}$ center is also likely to feature a dative bond with phosphorus.

While the electron distribution in P$_3$N$_5$ polymorphs is well-determined in the purely ionic approximation ($i.e.$ P$^{V+}$$_3$N$^{III-}$$_5$),[20] it is not as trivial in PN$_2$. The most likely configuration is [P$^{5+}$][N$_2$]$^{4-}$·e$^-$, in analogy with metallic binary subnitrides such as Ba$_2$N ([Ba$^{2+}$]$_2$[N$^{3-}$]·e$^-$).[51] This interpretation is strengthened by the bond length of the N$_2$ dimers, fitting that of a pernitride. The metallic character of PN$_2$ suggested from this analysis is compatible with the lack of a measurable Raman signal from the sample.

In order to analyze the stability of PN$_2$, variable cell structural relaxations at 137 GPa were performed. If atomic positions were allowed to freely relax, only a small difference in lattice constants of ~2% was found, but the distance between nitrogen atoms increased to ~1.9 Å compared to the experimentally-obtained ~1.42 Å (see Supporting Information). However, the relaxed structure features negative modes in its phonon dispersion relations calculated at T = 0 K in the harmonic approximation (Figure S11). In order to investigate the possibility of a temperature induced dynamical stabilization of PN$_2$, $ab$-$initio$ as well as classical molecular dynamics (MD) simulations based on a machine learned inter-atomic potential were performed (see Supporting Information). In both cases, no sign of a dynamical instability of PN$_2$ during the MD simulations at 300 K was observed. However, the N-N distances were strongly modified, independent of the starting configuration (all 1.4 Å or all 1.9 Å), and a mixture of ~37.5% ~1.4 Å ($i.e.$ single-bonded N-N) and ~62.5% ~2 Å N-N distances ($i.e.$ no N-N bonds) appeared (Figure S12 and Figure S13). Increasing the temperature to the experimental synthesis temperature of 2500 K led to an increased number of 1.4 Å N-N distances, $i.e.$ ~47% (Figure S14). This mixed bonding arrangement is, independent of temperature, overall stable over the full ~10 ps length of the $ab$-$initio$ MDs (92 atoms) and also stable up to 24 ns in the classical MDs (192 atoms), but N-N distances are found to switch between the two states dynamically.

To test whether this disordered theoretical model was a better fit to the experimental single-crystal XRD data, the data was additionally analyzed with the N1 position split into two (N1 and N1′) and their occupancy (sum constrained to 1) as well as coordinates refined independently. An occupancy of 92% and 8% was obtained for the N1 and N1′ atoms, with N1-N1 and N1′-N1′ distances of 1.3802(3) and 1.9820(3) Å, respectively. This disordered model has equal or slightly higher reliability factors (R-factors) than the original model, substantiating that the calculations capture the overall nature of the structure, showing a disorder in the N-N distances.

Additional theoretical calculations were performed to analyze the electronic configuration, including the assumed metallic character of the PN$_2$ solid. Calculations were performed for unit cell models without, with 25% and with 100% N-N single bonds. As seen in Figure S15, all models were found to have a non-zero DOS at the Fermi energy, demonstrating the metallicity of the PN$_2$ compound, in agreement with the electron distribution inferred from the compound's crystal chemistry. Computed ELFs, for example with 25% N-N bonds shown in Figure S16, nonetheless display the strong covalent nature of the P-N interaction, and the lack of lone electron pairs on the nitrogen atoms confirms the expected dative bonding between nitrogen and phosphorus.

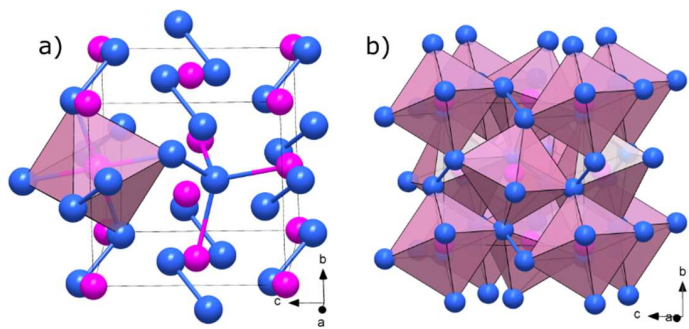

Figure 4: Crystal structure of PN$_2$ at 134 GPa, where the blue and pink spheres represent nitrogen and phosphorus atoms, respectively. a) Unit cell with a single PN$_6$ octahedra and the four bonds of the N[4] center drawn. b) Unit cell with all PN$_6$ octahedra shown.

Attempts to decompress PN$_2$ resulted in an additional datapoint at 109 GPa, below which the compound could no longer be observed. According to our calculations, PN$_2$ is predicted to have a bulk modulus of $K_0$ = 339.0 GPa ($V_0$ = 86.3 Å$^3$ and $K'$ = 4.21), therefore being even more incompressible than δ-P$_3$N$_5$ and qualifying it as an ultra-incompressible solid. This can be explained by the known incompressibility of compounds containing pernitride units, combined with the higher overall nitrogen coordination of PN$_2$ (*i.e.* only N[4] centers for PN$_2$ vs. a 3:2 mixture of N[4]/N[3] for δ-P$_3$N$_5$).

The synthesis of δ-P$_3$N$_5$ and PN$_2$, both containing PN$_6$ units, brings phosphorus nitrides—and pnictogen nitrides as a whole—in line with their periodic table neighbor silicon, germanium and tin nitrides (group 14 element nitrides) as well as sulfur nitrides (chalcogen nitrides), all of which feature XN$_6$ octahedra (X = Si, Ge, Sn and S).[15,16] While the Si-, Ge- and Sn-pernitrides, SiN$_2$, GeN$_2$, SnN$_2$, were all produced around 60 GPa,[15] the formation of the SN$_6$ units in SN$_2$ required a significantly higher pressure of 120 GPa.[16] In the case of PN$_6$, an intermediate pressure of 72 GPa was found sufficient. These formation pressures qualitatively match with the covalent radius (and electronegativity) of the non-nitrogen atom, with the larger elements (Sn, Ge, Si) requiring less pressure than the smaller ones (S)—according to the pressure homologue rule. It is interesting to note that pernitrides with the same formula type SiN$_2$, GeN$_2$, SnN$_2$, PN$_2$ and SN$_2$ are possible, and all but SN$_2$ adopt the pyrite-type structure with N[4] centers.[15] SN$_2$ has the CaCl$_2$-type structure with only N[3] centers,[16] and therefore a further increase in nitrogen coordination is expected to be achieved at higher pressures.

**Conclusions**

The results presented in this study, summarized in Figure 5, extend our understanding of the phosphorus-nitrogen system to a pressure of 137 GPa. Three previously unidentified nitride phases have been discovered, including two polymorphs of P$_3$N$_5$ (α′-P$_3$N$_5$ and δ-P$_3$N$_5$) as well as the pernitride PN$_2$. Most importantly, δ-P$_3$N$_5$ and PN$_2$ both feature the PN$_6$ unit, long sought-after in binary phosphorus nitrides, and bridge an important gap between group 14 element nitrides and chalcogen nitrides. Dative bonding is suggested from DFT calculations to enable the presence of N[4] centers in δ-P$_3$N$_5$ and PN$_2$—a first observation in phosphorus nitrides. The presence of both the PN$_6$ octahedra and the N[4] centers in δ-P$_3$N$_5$ and PN$_2$ provides a clear explanation for their very high incompressibility, respectively of 322(6) GPa and 339.0 GPa. δ-P$_3$N$_5$ is found non-recoverable to ambient conditions as it transforms into α′-P$_3$N$_5$ below 7 GPa. Although α′-P$_3$N$_5$ has the same PN$_4$ constituting units as α-P$_3$N$_5$, as well as the same atomic coordination, it is of significantly higher density (12.4%). As in the same way that α-P$_3$N$_5$ was considered for technological applications,[5,52] further studies on α′-P$_3$N$_5$ are necessary to assess its potential range of applicability. This investigation should stimulate further high-pressure high-temperature investigations on non-metal nitrides —largely neglected compared to metal nitrides—in order to gain a deeper understanding of their fundamental chemistry, which will contribute to the discovery of recoverable materials with uses in everyday life.

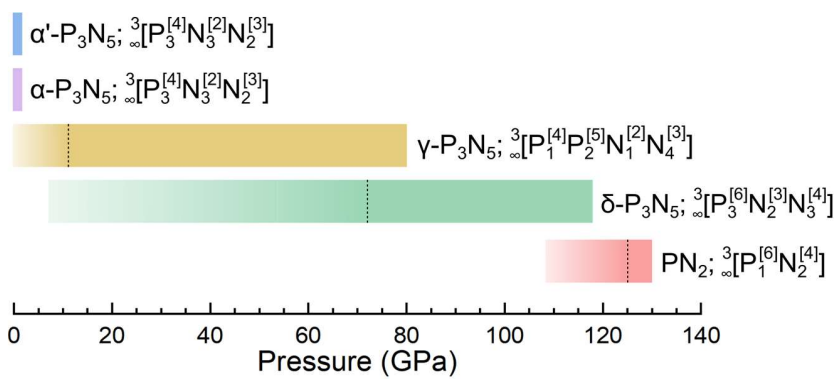

Figure 5: Schematic representation of the stability domains of the phosphorus-nitrogen system at high pressures. The colored horizontal lines show the pressure range at which a given phase was observed. The black vertical dashed line marks the lowest pressure at which a given compound was synthesized after laser-heating. Alongside each of the compounds' name, its Niggli representation is provided. α′- and α-P$_3$N$_5$ were solely observed at 1 bar.

**Methods**

*Experimental methods*

Three BX90-type screw-driven diamond anvil cell (DAC)[29] were prepared. One was equipped with 120 μm culet diamond anvils (120-DAC) and two others with 80 μm culet diamond anvils (80-DAC). Rhenium gaskets with an initial thickness of 200 μm were indented down to ~12-18 μm and sample cavities of about 40 to 60 μm in diameter were laser-drilled at the center of the indentations. One 80-DAC was loaded with red phosphorus (99.999%, Puratronics) while the two other DACs were loaded with black phosphorus. The procedure for producing black phosphorus is described in further detail below.[53] In both cases, the phosphorus piece was loaded alongside molecular nitrogen gas (~1200 bars), acting as a reagent and a pressure transmitting medium. The *in-situ* pressure was measured using the first-order Raman mode of the stressed diamond anvils[54] and verified through X-ray diffraction measurements by comparing to the pressure obtained from the calibrated equation of state of the rhenium gasket. Double-sided sample laser-heating was performed at our home laboratory at the Bayerisches Geoinstitut (BGI)[55] as well as at the GSECARS (APS) and P02.2 (PETRA III) beamlines with the phosphorus precursor employed as the laser absorber. Temperatures were measured with an accuracy of ±200 K, using the thermoemission produced by the laser-heated samples.[55]

Synchrotron X-ray diffraction measurements of the compressed samples were performed at ID11 (λ = 0.2852 Å) and ID15 (λ = 0.41015 Å) of the ESRF-EBS, at the P02.2 beamline (0.2910 Å) of DESY as well as at the GSECARS beamline (λ = 0.29521 Å) of the APS. In order to determine the position of the polycrystalline sample on which the single-crystal X-ray diffraction (SC-XRD) acquisition is obtained, a full X-ray diffraction mapping of the pressure chamber was achieved. The sample position displaying the most and the strongest single-crystal reflections belonging to the phase of interest was chosen for the collection of single-crystal data, collected in step-scans of 0.5° from −38° to +38°. The CrysAlisPro software[56] was utilized for the single crystal data analysis. The analysis procedure includes the peak search, the removal of the diamond anvils' and other 'parasitic' signal contributions, finding reflections belonging to a unique single crystal, the unit cell determination, and the data integration. The crystal structures were then solved and refined using the OLEX2 and JANA2006 software.[57,58] The SC-XRD data acquisition and analysis procedure was previously demonstrated and successfully employed.[16,59–61] The full details of the method can be found elsewhere.[62] Powder X-ray diffraction measurements were also performed to verify the sample's chemical homogeneity. The powder X-ray data was integrated with the Dioptas software.[63]

Confocal Raman spectroscopy measurements were performed with a LabRam spectrometer equipped with a ×50 Olympus objective. Sample excitation was accomplished using a continuous He-Ne laser (632.8 nm line) with a focused laser spot of about 2 μm in diameter. The Stokes Raman signal was collected in a backscattering geometry by a CCD coupled to an 1800 l/mm grating, allowing a spectral resolution of approximately 2 cm$^{-1}$. A laser power of about 4.6 mW incident on the DAC was employed.

Black phosphorus was synthesized out of red phosphorus using the 5000 tons uniaxial split sphere apparatus (Voggenreiter Zwick 5000[64]) at the BGI, under conditions of 2 GPa and 600°C maintained for 2 h. Lumps of red phosphorus (99.999%, Puratronics) were crushed in an agate mortar and subsequently loaded into a hexagonal boron nitride capsule to avoid any chemical reaction or oxidation during the synthesis. This capsule was then inserted into a 25/15 (octahedral edge length / anvil truncation edge length, in millimeter) BGI standard multi-anvil assembly equipped with a graphite heater. Temperature was monitored using a D-type thermocouple and kept constant during the time of the synthesis

*Density functional theory calculations*

Kohn-Sham density functional theory based structural relaxations and electronic structure calculations were performed with the QUANTUM ESPRESSO package[65–67] using the projector augmented wave approach.[68] We used the generalized gradient approximation by Perdew-Burke-Ernzerhof (PBE) for exchange and correlation, with the corresponding potential files: for P the 2p electrons and lower and for N the 1s electrons are treated as scalar-relativistic core states. We include van der Waals corrections following the approach by Grimme *et al.* as implemented in Quantum Espresso.[69] Convergence tests with a threshold of 1 meV per atom in energy and 1 meV/Å per atom for forces led to a Monkhorst-Pack[70] k-point grid of 8x16x8 for both α′-$P_3N_5$ and δ-$P_3N_5$ as well as $PN_2$ with a cutoff for the wavefunction expansion of 100 Ry for all phases. Phonon calculations were performed with PHONOPY[71] in 2x3x2 supercells for α′- and δ-$P_3N_5$ and 3x3x3 supercells for $PN_2$ with respectively adjusted k-points.

We performed variable cell relaxations (lattice parameters and atomic positions) on all experimental structures to optimize the atomic coordinates and the cell vectors until the total forces were smaller than $10^{-4}$ eV/Å per atom and the deviation from the experimental pressure was below 0.1 GPa

Furthermore, we calculated the equation of states (EOS) of both phosphorus nitrides by performing variable cell relaxations to respective target pressures until forces are < $10^{-3}$ eV/Å and until the pressure is matched within 0.1 GPa. A third order Birch Murnaghan (BM3) EOS was fitted to the calculated energy versus volume points. We obtained the following:

α′-$P_3N_5$:    $K_0$ = 95.4 GPa, $K'$ = 3.89, $V_0$ = 356.0 Å$^3$

δ-$P_3N_5$:    $K_0$ = 299.0 GPa, $K'$ = 4.21, $V_0$ = 125.1 Å$^3$

We benchmark the target pressure in the relaxations against the pressure obtained from the BM3 fit to ensure convergence. The EOS are in good agreement with experimental data obtained during compression and decompression (Figure 2).

In order to analyze the effect of temperature on the stability of the $PN_2$ phase and the N-N distances connecting the $PN_6$ octahedra, we ran *ab-initio* molecular dynamics (MD) simulations with Quantum Espresso using 2x2x2 supercells and 2x2x2 K-points and a timestep of 0.9697 fs. Simulations ran for 9.289 ps (300 K) and 9.869 ps (2500 K). We calculate the interatomic distance for each N-N pair as well as the percentage of short (< 1.6 Å) and long (> 1.6 Å) N-N distances in each timestep and obtain Figures S9 and S10. We find a stabilization of configurations with 37.5% short distances at 300 K and 47% short distances at 2500 K after ~3 ps and ~6 ps, respectively. To obtain a better understanding of the influence of cell size and simulation duration, we trained a moment tensor (mtp)[72] machine learning interatomic potential using the MLIP code[73] based on the structural relaxations and phonon calculations performed for $PN_2$, which was refined using an active learning process.[73] We trained the mtp potentials using *ab-initio* calculated total energies, interatomic forces and stresses of supercells with 192 atoms. For the *ab-initio* calculations in the active learning process, we used the *Vienna ab-initio simulation package* (VASP)[74–77] with the PBE generalized gradient approximation, 600 eV cutoff energy for the basis set and 4x4x4 sampling of the Brillouin zone. The final interatomic potential based on a 20g.mtp[72] was trained from 300 K up to 2500 K and 110 GPa up to 150 GPa. We evaluate the effect of enhanced cell size (up to 3000 atoms) and increased simulation time (up to 24 ns) at various temperatures and pressures through trajectory calculations with Lammps[78] using 1 fs timestep, calculating the time evolution of the radial distribution function (RDF). We find a very good agreement between the *ab-initio* MD and classical MD simulations: independently of the chosen cell size, simulation time and initial atomic arrangements the system evolves in a structural state with N-N distances of ~1.4 Å and ~2 Å (*c.f.* Figure S14).

Furthermore, we cannot find any influence greater than 2% on the N-N distances performing calculations using the Heyd–Scuseria–Ernzerhof hybrid functional[79] with the standard screening parameter, with and without van der Waals correction or spin polarization (magnetization along *z*-axis and noncollinear calculations with QE and VASP) starting from ferromagnetic and anti-ferromagnetic spin arrangements that could account for the remaining differences in interatomic N-N distances.

**Data Availability**

Full crystallographic data on the α′-$P_3N_5$, δ-$P_3N_5$ and $PN_2$ solids were deposited to the CCDC and can be accessed using the identifiers CSD 2178817-2178821. Other data that support the findings of this study are available from the corresponding author upon reasonable request.

**Acknowledgements**

The authors acknowledge the Advanced Photon Source (APS) and the European Synchrotron Radiation Facility (ESRF) for provision of beamtime at the GSECARS beamline and, the ID15b and ID11 beamlines, respectively. D.L. thanks the Deutsche Forschungsgemeinschaft (DFG, project LA-4916/1-1) and the UKRI Future Leaders Fellowship grant (MR/V025724/1) for financial support. N.D. and L.D. thank the Federal Ministry of Education and Research, Germany (BMBF, grant no. 05K19WC1) and the Deutsche Forschungsgemeinschaft (DFG projects DU 954–11/1, DU 393–9/2, and DU 393-13/1) for financial support. N.D. also thanks the Swedish Government Strategic Research Areas in Materials Science on Functional Materials at Linköping University (Faculty Grant SFO-Mat-LiU No. 2009 00971). I.A.A and F.T. are supported by the Swedish Research Council (VR) Grant No. 2019-05600. I.A.A. and Fe.T. acknowledge support from the VINN Excellence Center Functional Nanoscale Materials (FunMat-2) Grant 2016–05156. I.A.A. acknowledges also support from the Knut and Alice Wallenberg Foundation (Wallenberg Scholar grant no. KAW-2018.0194). Computations were enabled by resources provided by the Swedish National Infrastructure for Computing (SNIC) using Dardel at the PDC Center for High Performance Computing, KTH Royal Institute of Technology and LUMI at the IT Center for Science (CSC), Finland through grant SNIC 2022/6-10 and SNIC 2021/37-10, respectively. W.S. acknowledges funding support from the Deutsche Forschungsgemeinschaft (DFG, German Research Foundation) under Germany's Excellence Strategy-EXC 2089/1-390776260 (e-conversion). The authors also thank Florian Knoop for productive discussions. For the purpose of open access, the author has applied a Creative Commons Attribution (CC BY) licence to any Author Accepted Manuscript version arising from this submission.

**Keywords**